\documentclass[12pt]{iopart}
\pdfoutput=1
\usepackage{color}
\usepackage{amssymb}

\usepackage{graphicx} 
\usepackage{relsize}

\def\beqra{\begin{eqnarray}}
\def\eeqra{\end{eqnarray}}
\def\beq{\begin{equation}}
\def\eeq{\end{equation}}

\def\re#1{(\ref{#1})}

\def\agt{\stackrel{>}{\sim}}
\def\alt{\stackrel{<}{\sim}}
\begin{document}

\title{Spherical collapse and halo mass function in the symmetron model}
\author{Laura Taddei}
\address{Dipartimento di Fisica e Scienze della Terra, Universit\`a di Parma, Viale Usberti 7/A, I-43100 Parma, Italy.}
\ead{laura.taddei@pr.infn.it}
\author{Riccardo Catena}
\address{Institut f{\"u}r Theoretische Physik, Friedrich-Hund-Platz 1, D-37077 G{\"o}ttingen, Germany.}
\ead{catena@theorie.physik.uni-goettingen.de}
\author{Massimo Pietroni}
\address{INFN, Sezione di Padova, via Marzolo 8, I-35131, Padova, Italy.}
\address{Dipartimento di Fisica e Scienze della Terra, Universit\`a di Parma, Viale Usberti 7/A, I-43100 Parma, Italy.}
\ead{massimo.pietroni@pd.infn.it}

\begin{abstract}
We study the gravitational clustering of spherically symmetric overdensities and the statistics of the resulting dark matter halos in the ``symmetron model'', in which a new long range force is mediated by a $Z_2$ symmetric scalar field. Depending on the initial radius of the overdensity, we identify two distinct regimes: for small initial radii the symmetron mediated force affects the spherical collapse at all redshifts; for initial radii larger than some critical size this force vanishes before collapse because of the symmetron screening mechanism. As a consequence, halos with initial radii smaller than some critical value collapse earlier than in the $\Lambda$CDM and statistically tend to form more massive dark matter halos. Regarding the halo-mass function of these objects, we observe departures from standard $\Lambda$CDM predictions at the few percent level. The formalism developed here can be easily applied to other models where fifth-forces participate to the dynamics of the gravitational collapse. 
\end{abstract}

\section{Introduction}
Over the last decade, a wealth of evidence has been accumulated in favor of the 
conclusion that the expansion of our Universe is accelerating, mainly from the 
observation of  type-Ia supernovae \cite{Riess:1998cb, Perlmutter:1998np} and 
the cosmic microwave background radiation (CMB)~\cite{Ade:2013zuv} in 
combination with measurements of the Hubble constant and large-scale structures~\cite{Anderson:2013oza}. 

Assuming the validity of General Relativity (GR) on large scales, 
a possible explanation for the accelerated expansion is obtained by 
introducing a component of the cosmic fluid, the dark energy, with equation-of-state parameter $\omega<-1/3$.
The best fit model is currently very close to $\Lambda$CDM, which assumes 
that the dark energy is a cosmological constant, with equation-of-state parameter $\omega=-1$. Another 
possibility widely discussed in the literature is the quintessence, in which the dark 
energy has some dynamics, modeled by a scalar field rolling down a shallow 
potential~\cite{Ratra:1987rm,Frieman:1995pm,Caldwell:1998je}. For a generic potential, 
the requisite of shallowness implies that the excitations of the field are nearly massless, 
$m_{\phi}=\sqrt{V''\left(\phi\right)/2}\sim H_{0}\sim 10^{-33}$ eV.  If these light scalar 
fields exist, they should couple to the standard matter and hence introduce new observable long-range 
forces and time dependence of the constants of nature. As it has been discussed in the literature (starting from \cite{Carroll:1998bd}), a long-range fifth-force mediated by a nearly-massless scalar field coupled to the Standard Model via interactions with strength of order $1/M_{Pl}$ is incompatible with phenomenological constraints in the laboratory or in the solar system. Therefore, in order to be viable, the effect of these cosmological scalar fields should be screened in the local environment (for a review, see  \cite{Khoury:2010xi}).
 
An example of such {\em screening mechanisms} is at work in the {\em chameleon} models, discussed in \cite{Khoury:2003rn, 
 Brax:2004qh}.  In this scenario the matter-scalar coupling induces an environment-dependent mass for the scalar field, which becomes extremely massive when or where matter density is high. 
 
The {\em Vainshtein mechanism} 
 \cite{Vainshtein:1972sx,Deffayet:2001uk}, operates when the scalar has 
 derivative self-couplings which become important near matter sources such as 
 the Earth. The strong coupling essentially cranks up the kinetic 
 terms, which translates into a weakened matter coupling. Thus the scalar 
 screens itself and becomes invisible to experiments. This mechanism is central 
 to the phenomenological viability of braneworld modifications of gravity and 
 galileon scalar theories \cite{Dvali:2000hr,Nicolis:2008in}. 
 
The last mechanism, the one explored in this paper, is best known in the literature as the {\em symmetron mechanism}~\cite{Pietroni:2005pv,Olive:2007aj,Hinterbichler:2010es}. In its simplest implementation, a discrete symmetry is imposed on the scalar field. As a consequence, the matter-scalar coupling is non-vanishing only if the discrete symmetry is spontaneously broken, which happens when the environmental matter density drops below a critical value. 

As it was discussed in \cite{Brax:2011aw, Wang:2012kj}, the requirements imposed by local ({\it i.e.} solar system) GR tests imply that the scalar field has a cosmological range of less than $O(\mathrm{Mpc})$. Therefore, in these scenarios, no observable signature on linear and mildly non-linear scales (which are above $O(10\; \mathrm{Mpc})$) should be expected. Non-linear scales should then be attacked to work out the possible signatures of these models.
 
While $N-$body simulations have been performed for chameleon~\cite{Brax:2013mua}, Vainshtein~\cite{Li:2013nua}, and symmetron~\cite{Davis:2011pj,Brax:2012nk} models, a semi-analytic study is also welcome, in order to provide complementary physical insight on what is going on (see for instance, Ref.~\cite{Schmidt:2008tn} for a discussion of the chameleon mechanism in the context of $f(R)$ models and Ref.~\cite{Li:2011qda} for an extended excursion set approach to structure formation in chameleon models). In this paper, we study the symmetron model in the spherical collapse approximation, and then use the outputs of this analysis to compute the halo mass function and the linear bias. Contrary to the standard Einstein-de Sitter (EdS) or $\Lambda$CDM cases, the evolution of the spherical overdensity depends on its initial radius. Inside large overdensities the discrete symmetry is effectively restored long enough before collapse, so that the scalar field force is absent for most of the history of these objects, which basically follow the GR evolution. On the other hand, in small objects, the extra force is turned on all the way down to the collapse time. In order to simplify the computation we will consider the two extreme regimes of very large and very small objects, to get an idea of the maximal modification induced on relevant quantities such as the halo mass function. 
   
The paper is organized as follows:  our specific model for the symmetron scenario is introduced in Section 2, 
while  the spherically-symmetric solutions and the constraints from solar 
system tests of gravity are reviewed in Section 3.  
In Section 4 we review the spherical collapse model applied to $\Lambda$CDM
and EdS models and then apply it to the present scenario, both in the  \emph{thick-shell} and in the \emph{thin-shell} regimes. 
In Section 5, we use the spherical model results as inputs to compute the halo mass function and the linear bias. Finally, in Section 6, we give our conclusions and outline possible directions for future research.

\section{The model}
The symmetron model  \cite{Pietroni:2005pv, Olive:2007aj,  Hinterbichler:2010es} can be introduced as a particular scalar-tensor theory, described by the action,
\beq
\!\!\!\!\!\!\!\!\!\!\!\!\!\!\!\!S=\int d^4 x \sqrt{-g}\left[\frac{M_{Pl}^2}{2}\mathcal{R}-\frac{1}{2} \left(\partial\phi\right)^2-V\left(\phi^2 \right)\right] +\int d^4 x\sqrt{-\tilde{g}} \, \mathcal{L}_{m}\left(\psi,\tilde{g}_{\mu\nu}\right)\,,
\label{eq1}
\eeq
where
\begin{equation}
\tilde{g}_{\mu\nu} \equiv A^{2}\left(\phi^2\right)g_{\mu\nu}\,,
\label{eq2}
\end{equation}
and a $Z_2$ ($\phi \leftrightarrow -\phi$) symmetry is imposed on the two functions $V$ (scalar potential) and $A$ (conformal matter-scalar coupling), which therefore can depend only on integer powers of $\phi^2$.
${\cal R}$ is the Ricci scalar built from  $g_{\mu\nu}$ and $M_{Pl}\equiv ({\sqrt{8\pi G}})^{-1}$ where $G$ is the Newton's constant in the Einstein frame, $\mathcal{L}_{m}$ is the matter Lagrangian and Eq.~\re{eq2} relates the Einstein frame metric $g_{\mu\nu}$ to the Jordan frame one, $\tilde{g}_{\mu\nu}$. Since $\phi$ couples universally to all matter fields, the weak equivalence principle holds.
Varying the action with respect to the scalar field, we obtain the field equations for $\phi$:
\begin{equation}
\Box\phi-\frac{\partial V}{\partial\phi}-A^{3}\frac{\partial A}{\partial \phi}\tilde{T} = 0\,,
\label{eq3}
\end{equation}
where $\tilde{T}=\tilde{g}_{\mu\nu}\tilde{T}^{\mu\nu}$ is the trace of the Jordan frame energy-momentum tensor $\tilde{T}_{\mu\nu}=-\left(2/\sqrt{-\tilde{g}}\right)\delta\mathcal{L}_{m}/\delta \tilde{g}^{\mu\nu}$ which is covariantly conserved $\tilde{\nabla}_{\mu}\tilde{T}^{\mu}_{\nu}=0$.
For astrophysical objects, we may use the idealization of pressureless sources, so  $\tilde{T}\simeq-\tilde{\rho}$.
Written in terms of the density $\rho=A^{3}\tilde{\rho}$, which is conserved in the Einstein frame, the scalar field equation takes the form:
\begin{equation}
\Box \phi= \frac{\partial V}{\partial\phi}+\rho \frac{\partial A}{\partial \phi} \,.
\label{eq4}
\end{equation}
Therefore, the field evolves according to an effective potential
\begin{equation}
V_{eff}\left(\phi^2\right)=V\left(\phi^2\right)+\rho A\left(\phi^2\right)\,.
\label{eq5}
\end{equation}
We specify our model choosing the explicit forms for $V(\phi^2)$ and $A(\phi^2)$, 
\begin{equation}
V\left(\phi^2\right)=\bar{V}+V_{0}\,e^{-\frac{\phi^{2}}{2M^{2}}}\,,
\label{eq6}
\end{equation}
and
\begin{equation}
A\left(\phi^2\right)= e^{\frac{\lambda \phi^{2}}{2M^{2}}}\,,
\label{eq7}
\end{equation}
where $\bar{V}$ plays the role of a cosmological constant, $ V_{0} $ is a energy density which will turn out to be $\ll \bar V$,  $ \lambda $ is a dimensionless coupling constant (which will turn out to be $\ll 1$) and $M$ is a new mass scale \cite{Pietroni:2005pv}. Our results will not change qualitatively if a functional form different from exponential would be chosen for the functions  $V(\phi^2)$ and $A(\phi^2)$. Indeed, the phenomenological constraints reviewed in Sect.~3.2 imply that $\lambda,\;V_0/\bar V  < O(10^{-9})$ and $\phi/M =O(1)$, so we can safely expand $A$ to linear order in $\phi^2$ and consider a potential for  $V$ containing up to quadratic terms in $\phi^2$. In this case, a new parameter (the coefficient of $\phi^4$ in $V(\phi)$) would appear.

Assuming  $\lambda>0$, the effective potential $V_{eff}$ induces a density-dependent phase transition. Indeed, its second derivative in $\phi=0$ is given by
\beq
\left.\frac{d^2 V_{eff}}{d\phi^2}\right|_{\phi=0} = -V_0+\lambda \rho\,,
\eeq which  changes sign at a redshift $z_t$, given by.
\beq
\rho(z_t) = \rho_0(1+z_t)^3 = \frac{V_0}{\lambda}\,.
\label{zt}
\eeq 
For $z\ge z_t$ the minimum of $V_{eff}$ is at $\phi=0$, whereas for $z\le z_t$ two degenerate minima form at the $z$-dependent values
\beq
\phi_{min}(z) = \pm \,M\left[\frac{6 }{1+\lambda}\log\left(\frac{1+z_t}{1+z}\right)\right]^{1/2}\qquad\qquad(\mathrm{for}\;\;\;z\le z_t)\,.
\label{min}
\eeq
The coupling to matter is measured by the field-dependent quantity
\beq
\beta(\phi) = M_{Pl} \frac{d \log A\left( \phi^2\right) }{d\phi}=\lambda \frac{\phi M_{Pl}}{M^2}\,,
\label{betadef}
\eeq
which, if evaluated at the $z$-dependent minimum, vanishes for $z\ge z_t$. At $z=0$ it is given by
\begin{equation}
\beta_{0}\equiv \beta(\phi_{min}(z=0)) =\lambda  \frac{M_{Pl}}{M}\left[\frac{6}{1+\lambda}\log\left(1+z_{t}\right)\right]^{1/2}\,,
\label{eq14}
\end{equation}
where we have chosen the minimum with the $``+"$ sign in \re{min}. Notice that, with respect to the dark sector of the $\Lambda$CDM, the model presents three extra parameters: indeed, besides $\bar{V}$ and $\rho_0$, playing the roles of $\rho_\Lambda$ and $\rho_m$, respectively, we have the coupling $\lambda$,  the constant $V_0$, and the new mass scale $M$.  We decide to trade the latter for the more physically transparent parameters $z_t$, $\beta_0$, and $\mu\equiv M/M_{Pl}$. In the following, we will discuss the observational constraints on these parameters.

\section{Screening mechanism}
\subsection{Static solutions }
To study the constraints on the model from tests of gravity, we consider the symmetron profile around an astrophysical source. We model the latter by a sphere of radius $R$ and homogeneous mass density $\rho$, whereas the background energy density is given by $\bar{\rho}$.
The scalar field equation ($\ref{eq3}$) in spherical coordinates, and in the static limit,  reduces to
\begin{equation}
\frac{d^{2}\phi}{dr^{2}}+\frac{2}{r}\frac{d\phi}{dr}=V_{,\phi}+\rho A_{,\phi}\,.
\label{eq15}
\end{equation}
Analogously to what is done in \cite{Hinterbichler:2010es}, the radial field equation can be thought of as fictional particle rolling in a potential $-V_{eff}$, subject to the friction term $\frac{2}{r}\frac{d\phi}{dr}$. 
The solutions of the scalar field inside and outside the object were found in \cite{Hinterbichler:2010es}. They depend on a dimensionless parameter $\gamma$, called \emph{thin-shell} parameter, defined as
\begin{equation}
\gamma\equiv\frac{\lambda}{M^{2}}\left(\rho\left(t\right)-\bar{\rho}\left(t\right)\right)R^{2}=6\, \lambda\, \frac{M_{Pl}^{2}}{M^{2}}\Phi\,,
\label{eq37}
\end{equation}
where $\rho$ is the matter density inside the sphere, $\bar\rho$ is the cosmological one, and $\Phi$ the gravitational potential of the spherical overdensity with respect to the cosmological background.
Physically, this ratio measures the surface Newtonian potential relative to $M^2/ \lambda M_{Pl}^2$.
$\gamma$ will soon be interpreted as a \emph{thin-shell} factor for the solutions, in analogy with Chameleon models \cite{Khoury:2003rn}. Indeed, ($\ref{eq37}$) matches the chameleon \emph{thin-shell} expression, therefore, symmetrons and chameleons  have similar phenomenology, in particular for astrophysical tests.
If we rewrite the density inside the sphere as
\beq
\rho\left(t\right)=\rho_{i}\left(\frac{R_{i}}{R\left(t\right)}\right)^{3}\,, \nonumber
\eeq
where $\rho_{i}$ and $R_{i}$ are respectively the initial density and the initial radius of the sphere, and the density of the background as 
\beq
\bar{\rho}\left(t\right)=\bar{\rho}_{i}\left(\frac{a_{i}}{a\left(t\right)}\right)^{3}\,, \nonumber
\eeq
where $\bar{\rho}_{i}$ is the initial density of the background, the $\gamma$ parameter becomes:
\begin{equation}
\gamma\left(t\right)=\frac{\lambda}{\mu^{2}}3H_{0}^{2}R_{i}^{2}\left[\left(1+\delta_{m,i}\right)\left(\frac{R_{i}}{R\left(t\right)}\right)-\left(\frac{a_{i}}{a\left(t\right)}\right)^{3}\left(\frac{R\left(t\right)}{R_{i}}\right)^{2}\right]\,,
\label{eq39}
\end{equation}
where $ \delta_{m,i} $ is the initial density contrast, defined as $ \delta_{m,i}=\frac{\delta\rho_{i}}{\bar{\rho}_{i}}=\frac{\rho_{i}-\bar{\rho}_{i}}{\bar{\rho}_{i}}$.

Consider a test particle at a distance $R\ll r \ll m_{\phi}^{-1}$ away from the object, where $m_{\phi}$ is the mass of the scalar field. The scalar force to gravity ratio on this particle is \cite{Hinterbichler:2011ca}
\begin{equation}
\frac{F_{\phi}}{F_{N}}=-\frac{\beta\left(\phi\right)}{M_{Pl}}\frac{d\phi/dr}{F_{N}}\,,
\label{eq88}
\end{equation}
with $\beta\left(\phi\right)$ given in \re{betadef}.
Substituting the expression for the scalar field outside the object ($r>R$) into Eq.~($\ref{eq88}$), we have
\begin{equation}
\frac{F_{\phi}}{F_{N}}=6\frac{\beta\left(\phi\right)^{2}}{\gamma}\left[1-\sqrt{1/\gamma}\tanh\left(\sqrt{\gamma} \right) \right] \,,
\label{eq89}
\end{equation}
from which the dependence of the fifth force on the parameter $\gamma$ is manifest. Different astrophysical objects (stars, planets, galaxies) can be {\it screened} or {\it unscreened} according to their respective values for $\gamma$. If $\gamma \gg1$ Eq.~(\ref{eq89}) reduces to \cite{Hinterbichler:2011ca}
\begin{equation}
\frac{F_{\phi}}{F_{N}}\simeq6\frac{\beta\left(\phi\right) ^{2}}{\gamma}\ll1\,,
\end{equation}
and the object is screened.
In this regime the field inside the object  is exponentially suppressed with respect to the asymptotical value outside, except within a \emph{thin-shell} beneath the surface.
In the opposite regime,  $\gamma\ll1$, we can Taylor expand  Eq.~(\ref{eq89}), which gives \cite{Hinterbichler:2011ca}
\begin{equation}
\frac{F_{\phi}}{F_{N}}\simeq2\beta^{2} \,.
\label{eq41}
\end{equation}
There is no \emph{thin-shell} in this case; the scalar field has basically the same value inside and outside the object, hence the symmetron couples with gravitational strength to the entire source.

\subsection{Constraints from Tests of Gravity}
Since the field is long ranged (and universally coupled) in almost all situations today the theory is best constrained by solar system experiments which have been performed with high precision. In this subsection, we adapt the findings of~\cite{Hinterbichler:2010es} to the present implementation of the symmetron scenario.
Requiring that our Galaxy is sufficiently screened, namely, that  $ \gamma_{G} > 10$, gives (from Eq.~\re{eq37} and using   $\Phi_{G}\sim 10^{-6}$) 
\begin{equation}
\frac{M}{M_{Pl}}=\mu < \sqrt{\lambda\, \Phi_{G}} \lesssim 10^{-3} \lambda^{1/2}\,.
\label{eq42}
\end{equation}
In this parameter regime, the Sun $ \left( \Phi_{\odot}\sim 10^{-6} \right)$ is  also screened, but the Earth  $ \left( \Phi_{\oplus}\sim 10^{-9} \right)$ is not \cite{Hinterbichler:2010es}.

GR tests in the solar system give constraints  on the two  post-Newtonian parameters, $\gamma_{PPN}$ and $\beta_{PPN}$ \cite{Will:2001mx}, which can be expressed in terms of the scalar coupling $\beta(\phi)$ of Eq.~\re{betadef}.
The tightest constraint on $\gamma_{PPN}$ comes from time-delay and light-deflection observations. In the present model, they imply
\begin{equation}
|\gamma_{PPN}-1 | = 2\frac{\beta(\phi)^2}{1+\beta(\phi)^2}\approx 2 \lambda^2 \phi^2 \frac{M_{Pl}^2}{M^{4}}= \frac{1}{3}\gamma \frac{\phi^2}{M^2}\,\frac{ \lambda}{\Phi} \,,
\label{gammappn}
\end{equation}
where, to obtain the last equality, we have used Eq.~\re{eq37}.
Near the Sun, using the solution of the field equation for the screened case \cite{Hinterbichler:2010es}, we have that $\phi=\phi_{\odot}\approx \phi_{G}/\sqrt{\gamma_{\odot}}$ where $\phi_{G}$ is the asymptotic value of $\phi$ inside the galaxy.   The relation between $\phi_{G}$ and the asymptotic cosmological value today, $\bar\phi_0$, is also obtained from the solution of the field equation,
\beq
\sqrt{\gamma_\odot} \frac{\phi_\odot}{M}\simeq \frac{\phi_G}{M}\simeq\frac{\bar\phi_0}{M} \frac{R_G}{R_{s.s.}} \frac{1}{\sqrt{\gamma_G}} \e^{\sqrt{\gamma_G}\left(1-\frac{R_{s.s}}{R_G}\right)}\simeq3\times 10^{-2}\, \frac{\bar\phi_0}{M} \,,
\label{sol2}
\eeq
where $R_G\sim100\,\mathrm{kpc}$ is the galactic radius,  $R_{s.s.}\sim10\,\mathrm{kpc}$ is the distance between the solar system and the galactic center and, following  \cite{Hinterbichler:2010es}, we have adopted  the fiducial value  $\gamma_G=20$. Inserting \re{sol2} in \re{gammappn} we have 
\beq
|\gamma_{PPN}-1 | \simeq 3\times 10^{4} \lambda \left(\frac{\bar\phi_0}{M}\right)^2\,,
\eeq
where we have also used $\Phi_{\odot}\sim 10^{-6} $.
Since from \re{min} we have $\bar\phi_0/M=O(1)$, the  current constraints from the Cassini spacecraft \cite{Bertotti:2003rm}, $|\gamma_{PPN}-1|\approx 10^{-5}$, can be satisfied for $\lambda \alt 10^{-9}$. Similar bounds come from the 
Nordvedt Effect, which describes the difference in free-fall acceleration of the Moon and the Earth towards the Sun due to scalar-induced differences in their gravitational binding energy~\cite{Williams:2003wu}. 

Finally, constraints from binary pulsars are trivially satisfied, since both the neutron star and its companion are screened. As we can see from \cite{Hinterbichler:2010es}, the force between these bodies is therefore suppressed by two \emph{thin-shell} factors:
\begin{equation}
\frac{F_{\phi}}{F_{N}}=\frac{1}{\gamma_{pulsar}}\cdot\frac{1}{\gamma_{companion}} \,.
\end{equation}
Estimating $\Phi_{pulsar}\sim 0.1$ and $\Phi_{companion}\sim 10^{-6}$, then for our fiducial parameter choices we obtain $\gamma_{pulsar}\sim 10^{5}$, $\gamma_{companion}\sim 10$ and therefore $\frac{F_{\phi}}{F_{N}}\approx 10^{-6}$, well below the current pulsar constraints on scalar-tensor theories.

The scalar field mass in the cosmological background is given by the second derivative of the effective potential \re{eq5} evaluated in $\bar\phi_0$. It is
\beq
m^2_\phi\sim \frac{V_0}{M^2} \sim \frac{\lambda}{\mu^2}\frac{\rho_0}{M_{Pl}^2} >10^6 H_0^2\,, 
\eeq
where we have used Eqs.~\re{zt} and \re{eq42}. This is in agreement with the findings of Refs.~\cite{Brax:2011aw, Wang:2012kj} and implies that the scalar field range is smaller than $O(\mathrm{Mpc})$, and therefore the scalar force gives no observable signature on linear and mildly non-linear scales, which are above $O(10\; \mathrm{Mpc})$. In the next sections we will discuss the effect of the symmetron model on non-linear scales, by using the spherical collapse approximation. 

\section{Spherical collapse}
A standard approach to follow the evolution of cold dark matter structures during the first stages of the non-linear regime is the spherical collapse model \cite{Gunn:1972sv,Padmanabhan,Peacock}. This approach was first applied to the EdS Universe and later on in the context of the $\Lambda$CDM~\cite{Lahav:1991wc}. Recently, the spherical collapse approximation has been also extended to quintessence models, as for instance in~\cite{Pace:2010sn,Wintergerst:2010ui}. In the following we will briefly review the basic equations in the EdS and $\Lambda$CDM cases and then we will extend these to the symmetron model.

\subsection{Application to standard cosmologies}
Consider a spherical density perturbation of radius $R$ within a homogeneous background Universe. Under the effect of the gravitational attraction, the perturbation grows, possibly entering the nonlinear regime, depending on the scale of the perturbation. As a consequence of Birkhoff's theorem we can treat the spherical overdensity as a closed Universe where the total density $\rho=\bar{\rho}+\delta\rho$ exceeds the density of the background $\bar{\rho}$ due to the presence of the density perturbation. The radius $ R $ evolves according to the Friedmann equation:
\begin{equation}
\frac{\ddot{R}}{R}=-\frac{1}{6M_{Pl}^{2}}\sum_{\alpha}\rho_{\alpha}\left[ \left(1+3w_{\alpha}\right)\right]  
\label{eq60}
\end{equation}
where the sum is over particle species. This sphere is embedded in a homogeneous Friedmann-Robertson-Walker (FRW) background characterized by the scale factor $a\left( t\right) $ and the Hubble function $ \bar{H}\equiv\dot{a}/a $. We use a bar to indicate background quantities. With this notation the Friedmann equations describing the homogeneous and flat background Universe are:
\beqra
\label{eq:Hbar}
\bar{H}^{2} &=& \frac{1}{3M_{Pl}^{2}}\sum_{\alpha}\bar{\rho}_{\alpha} \\
\label{eq33}
\frac{\ddot{a}}{a} &=& -\frac{1}{6M_{Pl}^{2}}\sum_{\alpha}\bar{\rho}_{\alpha}\left[ \left(1+3\bar{w}_{\alpha}\right)\right] \,.
\eeqra
In the EdS and $\Lambda$CDM scenarios, the matter energy density $\rho_{m}$ and the cosmological constant energy density $\rho_{\Lambda}$ are conserved, both inside and outside the spherical perturbation:
\beqra
&&\dot{\rho}_{m}+3H\rho_{m}=0 \label{eq34} \\
&&\dot{\bar{\rho}}_{m}+3\bar{H}\bar{\rho}_{m}=0 \label{eq35} \\
&&\bar{\rho}_{\Lambda}=\rho_{\Lambda}= \rm const.
\label{eq36}
\eeqra
The non-linear density contrast is defined by $1+\delta_{m}\equiv \rho_{m}/\bar{\rho}_{m}$ and it is determined by Eqs.~($\ref{eq34}$) and  ($\ref{eq35}$). Linear perturbation theory \cite{Ma:1995ey}, on the other hand, gives the evolution equation:
\begin{equation}
\ddot{\delta}_{m,L}+2\bar{H}\dot{\delta}_{m,L}-\frac{1}{2M_{Pl}^{2}}\bar{\rho}_{m}\delta_{m,L}=0 \,.
\label{eq38}
\end{equation}
Eqs.~($\ref{eq60}$)--($\ref{eq38}$) can be integrated numerically. We start the integration at some initial time $t_{in}$ when the total energy density in the spherical overdensity is higher than the critical energy density, due to the presence of the perturbation $\delta_{m} $. Eq.~($\ref{eq60}$) gives the function $R\left(z\right)$. This first increases as the spherical perturbation expands following the background evolution, then it reaches a maximum value (turnaround) in which comoving velocities become zero; finally, the sphere collapses and its radius tends to zero. 

The redshift of collapse depends on the initial density contrast $\delta_{m,i}$: the higher is $\delta_{m,i}$, the earlier the overdense region collapses. The corresponding value of the linear density contrast extrapolated at the time of collapse is referred to as $\delta_{c} $ and it can be calculated by stopping the evolution of Eq.~($\ref{eq38}$) when $\delta_{m}$ goes to infinity, i.e. the overdensity collapses. Varying the initial conditions one obtains different collapse redshifts $z_{c}$. In this way we derive the redshift dependence of the critical density $\delta_{c}=\delta_{c}\left( z_{c}\right)$. To be sure of starting the integration when overdensities are still linear, we find that it is necessary to work in a range of initial overdensities with $\delta_{m,in}\ll 1$ (in our numerical computations, we take $\delta_{m,in}\alt 10^{-3}$) . The quantity $\delta_{c}$ is important because it represents one of the key ingredients to calculate the halo-mass function, which provides a statistical information on the mass distribution of the collapsed spherical overdensities, {\it i.e.} the dark matter halos (see Sec. 5). 

\begin{figure}[t]
\centerline{\includegraphics[width = 10cm,keepaspectratio=true]{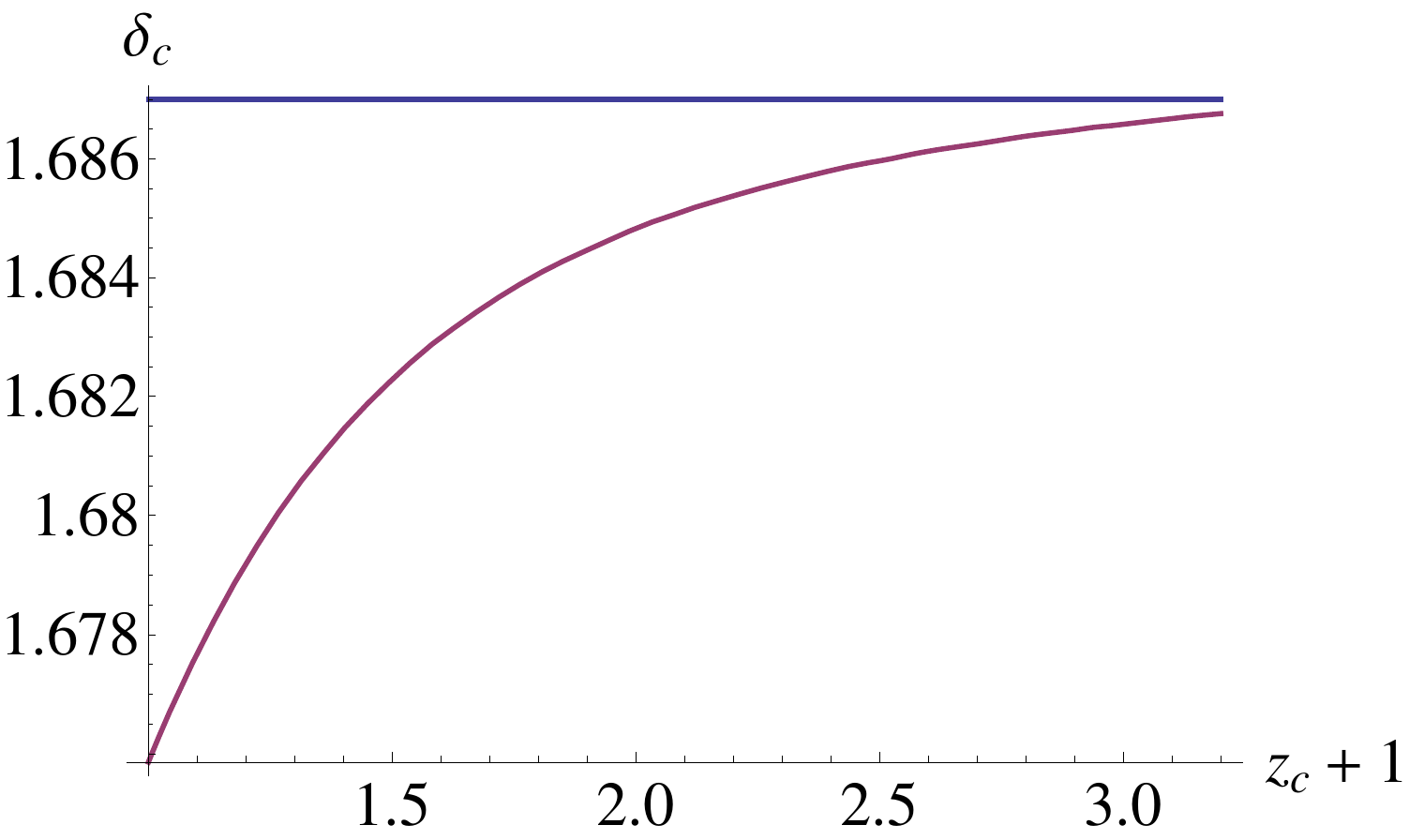}}
\caption{Extrapolated linear density contrast at collapse $ \delta_{c} $ vs $z_{c}+1$ for EdS (blue curve) and $\Lambda$ CDM (red curve) models.}
\label{Fig2}
\end{figure}

In an EdS scenario the linear density contrast at collapse can be calculated analytically~\cite{Padmanabhan,Peacock} and it is equal to a constant value independent of the redshift of collapse $z_{c}$: 
\begin{equation}
\delta_{c}=\left( 3/20\right) \left( 12 \pi\right) ^{2/3}\simeq 1.686 \,.
\end{equation}
In the $\Lambda$CDM case, instead, this value decreases for late collapse times, when dark energy dominates over matter and leads to cosmic acceleration, slowing down the structure formation. This well-known effect is shown in Fig.~\ref{Fig2}, where we plot $\delta_{c}\left( z_{c}\right)$ for the EdS and the $\Lambda$CDM scenarios.

\subsection{Application to the symmetron model}
Now we generalize the spherical collapse equations to the case of the symmetron model. We make here a few simplifying assumptions which nevertheless allow us to properly account for the main features of the symmetron mediated force. 
We approximate the coupling function $\beta(\phi)$ inside the spherical overdensity to be equal to the cosmological one,  $\beta(\phi_{min}(z))$, as long as the parameter $\gamma$, computed according to Eq.~\re{eq39}, is smaller than a threshold value $\gamma_{tr}=\mathcal{O}(1)$, and to drop to zero everywhere inside the overdensity as soon as $\gamma> \gamma_{tr}$ (for a refined treatment of the scale dependence of the coupling $\beta(\phi)$ and of the overdensity profile, see Ref.~\cite{Brax:2013fna}). Moreover, we consider only spherical overdensities characterized by $m_{\phi}R <1$, a requirement which considerably simplifies the Poisson equation for the scalar field fluctuations. Within these assumptions we can treat the evolution of the sphere of radius $R$ as the one of a closed Universe coupled to the background expansion. 

We start with a qualitative description of the spherical collapse in the symmetron model. In general, before the phase transition $ \left(z > z_{t} \right)$ the symmetron sits on the minimum $\phi=\phi_{min}=0$ and the model coincides with the $\Lambda$CDM. After the phase transition $\left(z < z_{t} \right)$, the symmetron evolves towards a different minimum $\phi=\phi_{min}\neq0$ \re{min} which implies $\beta(\phi)\neq0$ and therefore a spherical collapse potentially different from the one of the $\Lambda$CDM. Let us now assume $\beta(\phi)\neq0$. Initially, for all the time before the phase transition, the scalar field is zero inside the sphere and in the background, so it is not important to clarify if the sphere is in the thick-shell or in the thin-shell regime; only after the phase transition, the subsequent evolution of the sphere will always drive the $\gamma$ parameter towards values larger than the $\mathcal{O}(1)$ threshold $\gamma_{tr}$, independently from the initial radius $R_i$ of the overdensity, as one can see taking the limit $R\rightarrow 0$ in Eq.~(\ref{eq37}). In practice, however, dark matter halos form at the virialization radius $R=R_{vir}$, corresponding to a value of $\gamma$ which we denote here by $\gamma_{vir}$. This allows us to clearly identify two different scenarios: for small enough initial radii the overdensity virializes when $\gamma_{vir}<\gamma_{tr}$, and therefore the sphere remains in the thick-shell regime until the associated dark matter halo has formed. For large initial radii, instead, dark matter halos form when $\gamma_{vir}>\gamma_{tr}$, which implies a transition from the thick-shell regime to the \emph{thin-shell} regime, defined in fact by the condition $\gamma>\gamma_{tr}$.

We now introduce the equations which quantitatively describe the spherical collapse in the symmetron model. We denote by $\phi$ the scalar field inside the sphere and by $\bar{\phi}$ the background scalar field. The flat background Universe is described by the Friedmann equations:
\beqra
\label{eq30}
&&\frac{\ddot{a}}{a}=-\frac{1}{6M_{Pl}^{2}}\left[\bar{\rho}_{m}A\left(\bar{\phi}^2\right)+\bar{\rho}_{\phi}+3\bar{p}_{\phi}\right] \\
\label{eq31}
&&\left(\frac{\dot{a}}{a}\right)^{2}=\frac{1}{3M_{Pl}^{2}}\left[\bar{\rho}_{m}A\left(\bar{\phi}^2\right)+\bar{\rho}_{\phi}\right]
\eeqra
where
\begin{equation}
\bar{\rho}_{\phi}=\frac{1}{2}\dot{\bar{\phi}}^{2}+V\left(\bar{\phi}^2\right)
\label{eq45}
\end{equation}
\begin{equation}
\bar{p}_{\phi}=\frac{1}{2}\dot{\bar{\phi}}^{2}-V\left(\bar{\phi}^2\right)
\label{p}
\end{equation}
are respectively the density and the pressure of the background scalar field. We can derive the evolution equation for the sphere radius $R$ by following the same steps of \cite{Wintergerst:2010ui}. We find:
\begin{equation}
\frac{\ddot{R}}{R} =-\beta\left(\phi\right)\dot{\bar{\phi}}\left( \bar{H}-\frac{\dot{R}}{R}\right)+\frac{\ddot{a}}{a}-\frac{1}{6M_{Pl}^{2}}\bar{\rho}_{m}\delta_{m}\left[1+2\beta\left(\phi\right)^{2}\right] \,.
\label{eq49}
\end{equation}
This equation describes the general evolution of the radius of a spherical overdense region subject to a scalar coupling $\beta(\phi)$, which controls the terms responsible for the additional attractive force. Since the scalar field slowly evolves following the background minimum $\phi_{min}(z)$, we can safely neglect all terms proportional to $\dot{\bar{\phi}}$ in Eqs.~(\ref{eq30}), (\ref{eq31}) and (\ref{eq49}). This leads to
\beqra
\label{eqa}
&&\frac{\ddot{a}}{a}=-\frac{1}{6M_{Pl}^{2}}\left[\bar{\rho}_{m}A\left(\bar{\phi}^2\right)-2V\left(\bar{\phi}^2\right)\right] \\
\label{eqb}
&&\left(\frac{\dot{a}}{a}\right)^{2}=\frac{1}{3M_{Pl}^{2}}\left[\bar{\rho}_{m}A\left(\bar{\phi}^2\right)+V\left(\bar{\phi}^2\right)\right] \\
\label{eqc}
&&\frac{\ddot{R}}{R} =\frac{\ddot{a}}{a}-\frac{1}{6M_{Pl}^{2}}\bar{\rho}_{m}\delta_{m}\left[1+2\beta(\phi)^{2}\right] \,.
\eeqra

\begin{figure}[t]
\centerline{\includegraphics[width = 12cm,keepaspectratio=true]{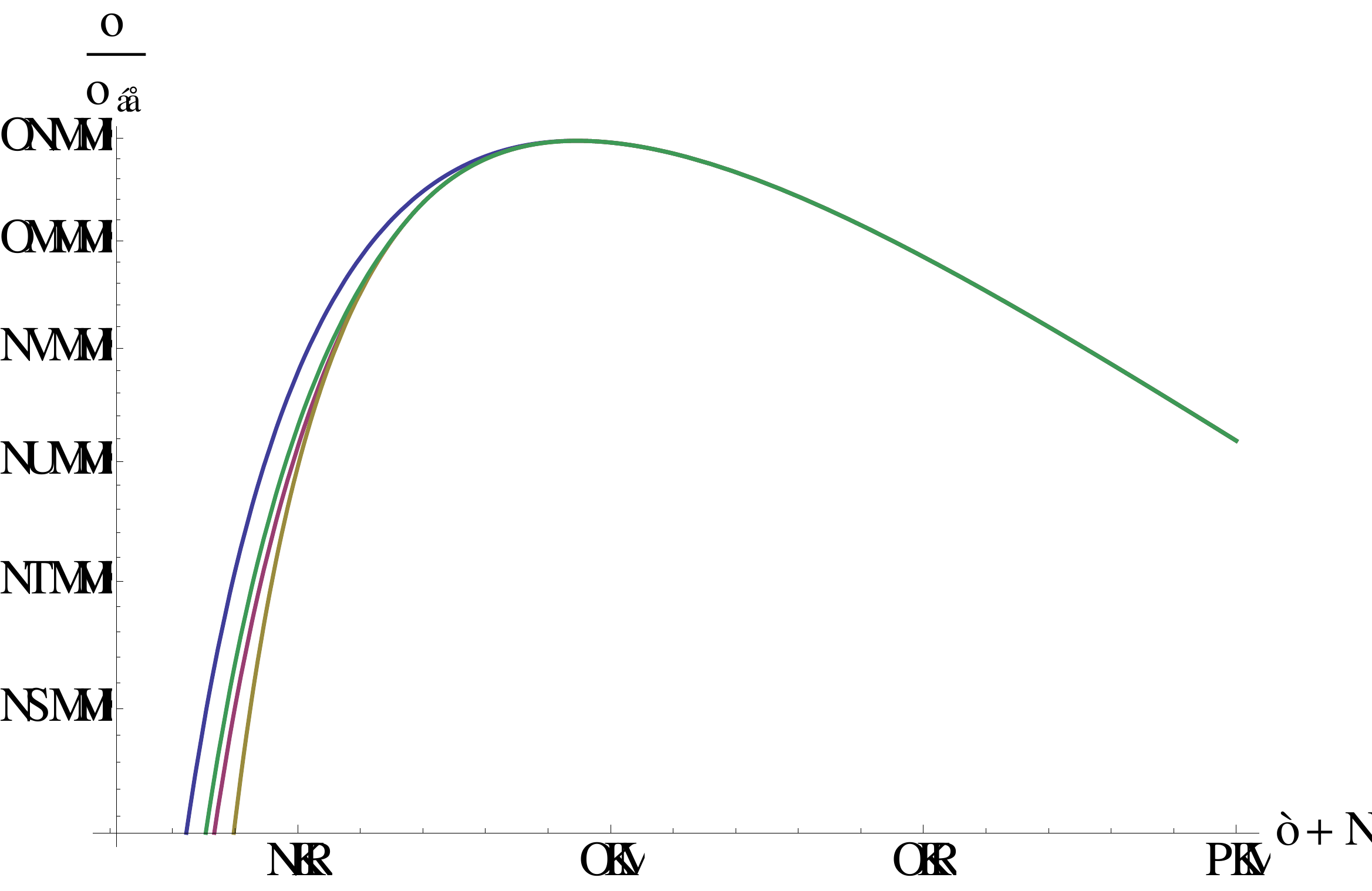}}
\caption{Evolution of radius $\frac{R}{R_{i}}$ vs  redshift for $\delta_{m,i}=0.0003$. The blue curve represents the $\Lambda$ CDM model, the green curve and the red curve are associated with the transition regime from \emph{thick} to \emph{thin-shell} obtained by fixing $R_{i} \sim 4$ Mpc h$^{-1}$ and $R_{i} \sim 3.5$ Mpc h$^{-1}$ respectively and the yellow curve represents the \emph{thick-shell} regime, obtained by fixing $R_{i} = 0.9$ Mpc h$^{-1}$. As we can see, before the phase transition ($z\geq1$) the models coincide with the $\Lambda$ CDM. After the phase transition ($z<1$), we observe that the sphere collapses earlier in the \emph{thick-shell} regime  (yellow curve) with respect to the other models. We can also see that at $z \sim 0.6$ the yellow curve starts to be different from the red curve: this point represents the transition from \emph{thick} to \emph{thin-shell} and the moment in which $\gamma $ becomes larger than $\gamma_{tr}$.}
\label{Fig1}
\end{figure}
We now present the results of a numerical integration of the equations for the spherical collapse in the symmetron model. In our examples, we will fix the scalar coupling today, defined in \re{eq14}, at $\beta_0=1$, and we fix $z_t=1$ and $\mu=10^{-9}$.
With these parameters, spherical overdensities  of initial radii smaller than $~0.9 \;\mathrm{Mpc\,h^{-1}}$ in comoving units do not cross the  $\gamma = 1$ threshold before virialization, therefore their evolution takes place entirely in the thick shell regime.
If we integrate numerically Eqs.~(\ref{eqa}), (\ref{eqb}) and (\ref{eqc}) from $z_{i}\sim7000$ to $z_{f}=0$, with an initial density contrast $\delta_{m,i} = 0.0003$ and setting $\phi(z)=\phi_{min}(z)$ at all $z$, we observe that the radius collapses at $z \simeq 0.3$, as we can see from Fig.~\ref{Fig1}. 

For  large initial radii, namely $R_{i} \agt  0.9 \;\mathrm{Mpc\,h^{-1}}$,  the spherical collapse passes through a thick-shell/thin-shell transition before collapsing.  After the transition the scalar force is confined only within a thin-shell beneath the surface so the sphere collapses later with respect to the case in which the object is unscreened, as one can see from Fig.~\ref{Fig1}.

\begin{figure}[t]
\centerline{\includegraphics[width = 12cm,keepaspectratio=true]{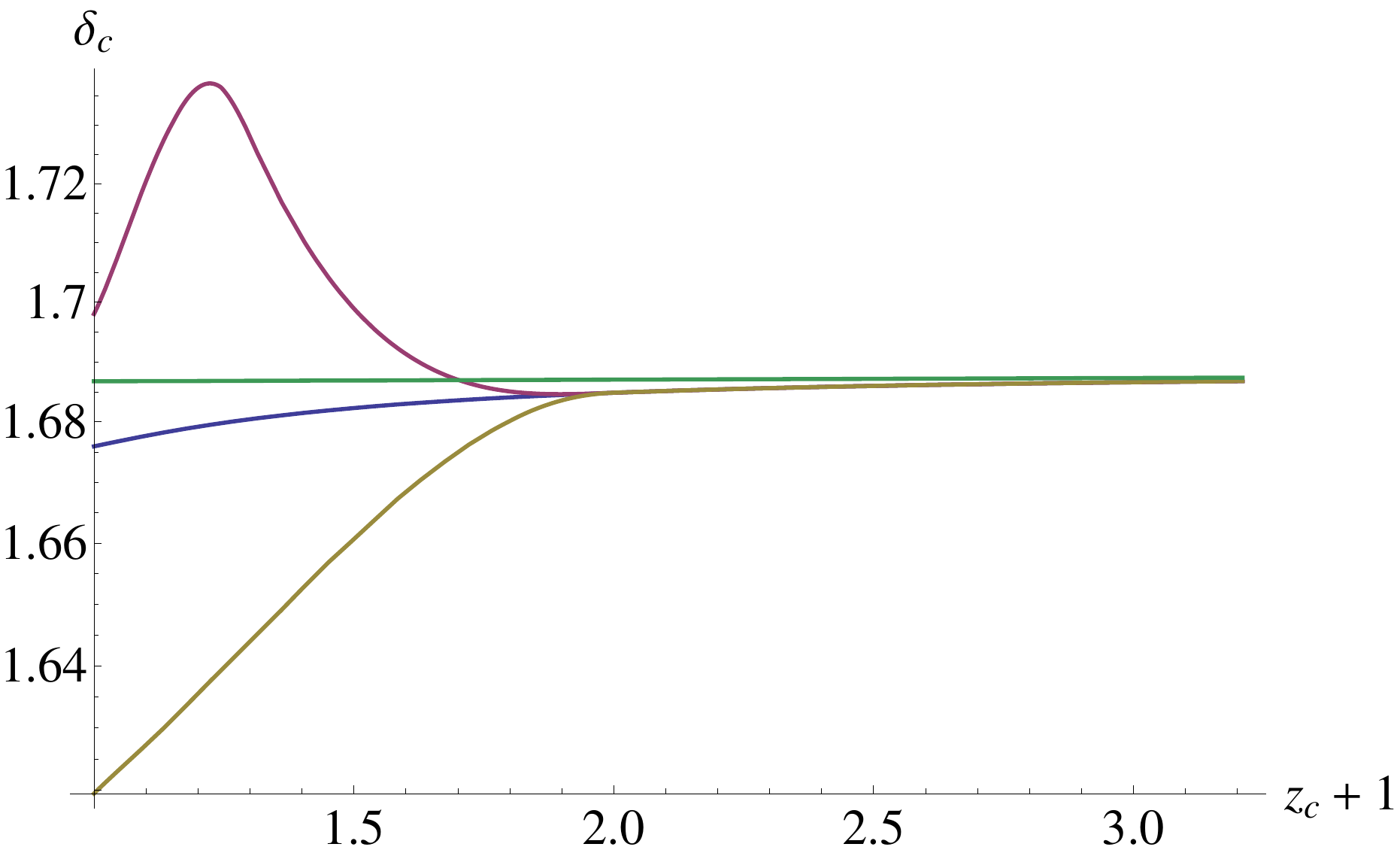}}
\caption{Extrapolated linear density contrast at collapse $ \delta_{c} $ vs $z_{c}+1$ for EdS (green curve), $\Lambda$ CDM (blue curve), transition from \emph{thick} to \emph{thin-shell} regime (red curve) and \emph{thick-shell} regime (yellow curve) models.}
\label{Fig3}
\end{figure}
Having determined the evolution of $R$ in time, we can now determine $\delta_c$. We follow the same steps of \cite{Wintergerst:2010ui} to obtain the non-linear evolution of the density contrast:
\begin{equation}
\ddot{\delta}_{m}=\left(\beta\dot{\bar{\phi}}-2\bar{H}\right)\dot{\delta}_{m}+\frac{4}{3}\frac{\dot{\delta}_{m}^{2}}{1+\delta_{m}}+\frac{1+\delta_{m}}{a^{2}}\nabla^{2}\Phi_{eff} \,.
\end{equation}
Linearization leads to~\cite{Brax:2011pk,Amendola:2003wa,Amendola:1999dr,Pettorino:2008ez}
\begin{equation}
\ddot{\delta}_{m,L}=\left(\beta\dot{\bar{\phi}}-2\bar{H}\right)\dot{\delta}_{m,L}+\frac{1}{a^{2}}\nabla^{2}\Phi_{eff}
\label{eq62}
\end{equation}
where $\Phi_{eff} $ is the effective gravitational potential given by
\begin{equation}
\Phi_{eff}\equiv\Phi+\beta\delta\phi
\label{eq63}
\end{equation}
which obeys the modified Poisson equation
\begin{equation}
\nabla^{2}\Phi_{eff}=\frac{a^{2}}{2M_{Pl}^{2}}\bar{\rho}_{m}\delta_{m}\left(1+2\beta^{2}\right) 
\label{eq64}
\end{equation}
and $\Phi$ the usual gravitational potential. Since the scalar field is slowly varying during the spherical collapse, we can neglect all terms proportional to $\dot{\bar{\phi}}$ in Eq.~(\ref{eq62}) and write 
\begin{equation}
\ddot{\delta}_{m,L}\simeq-2\bar{H}\dot{\delta}_{m,L}+\frac{1}{a^{2}}\nabla^{2}\Phi_{eff}
\label{eq65}
\end{equation}
where $\bar{H} $ is defined in Eq.~(\ref{eq:Hbar}). When $\beta\left(\phi\right)$=0, Eq.~(\ref{eq65}) coincides with Eq.~(\ref{eq38}). We numerically solve Eq.~(\ref{eq65}) from $z_{i}\sim7000$ to $ z_{f}=0 $ and calculate the value of the linear density contrast at collapse $\delta_{m,L}\left(z=z_{c}\right)$ for different $z_c$ by varying the value of $\delta_{m,i} $, as we have done for the $\Lambda$CDM case. From Fig.~\ref{Fig3} one can see that the two scenarios studied here (\emph{thick-shell} regime and transition from \emph{thick} to \emph{thin-shell} regime) approach the $\Lambda$CDM prediction at high redshifts corresponding to the $Z_2$-symmetric phase. At $z=0$, the difference in $\delta_{c} $ with respect to the $\Lambda$CDM case is about 2\% for the \emph{thin-shell} regime and about 5\% for the \emph{thick-shell} regime, with the present choice of parameters. 

\section{Halo mass function and bias}
The halo-mass function is defined as the comoving number density of halos per logarithmic interval in the virial mass $M_v$~\cite{Padmanabhan,Peacock}. Recently there has been a renewed theoretical interest regarding the halo mass function which has led to a variety of novel approaches to its determination, including interesting applications of the path integral formalism~\cite{Maggiore:2009rv}. In the present analysis, which aims at comparing the predictions of our symmetron model with standard $\Lambda$CDM results, we will use a simple prescription for the halo mass function which relies on the Press-Schechter theory and on a scaling function first proposed by Sheth and Tormen (ST)~\cite{Sheth:1999mn}. This approach has been tested in various frameworks (see for instance~\cite{Schmidt:2008tn}) and it guarantees sufficient accuracy for the purposes of the present work. Within these assumptions, the halo mass function takes the following form~\cite{Sheth:1999mn}
\begin{equation}
n_{\ln M_{v}}\equiv\frac{dn}{d\ln M_{v}}=\frac{\bar{\rho}_{m}A(\bar{\phi})}{M_{v}}f\left(\nu\right)\frac{d\nu}{d\ln M_{v}} \,,
\label{mf}
\end{equation}
where the linear power spectrum entering in the variable $\nu=\delta_{c}/\sigma\left(M_{v}\right)$ is evaluated at the present time. Here $\sigma\left(M_v\right)$ is the variance of the linear density field convolved with a top hat of radius $r$ that encloses the mass $M_v=4\pi r^{3}\bar{\rho}_{m}A(\bar{\phi})/3$, namely
\begin{equation}
\sigma^{2}\left(M\right)=\int\frac{d^{3}k}{\left(2\pi\right)^{3}}\left|\tilde{W}\left(kr\right)\right|^{2}P_{L}\left(k\right) \,,
\label{sigma}
\end{equation}
where $P_{L}\left(k\right)$ is the linear power spectrum (computed with the same linear growth equations used in the computation of $\delta_c$)  and $\tilde{W}$ is the Fourier transform of the top hat window function. The ST scaling function appearing in Eq.~(\ref{mf}) is given by
\begin{equation}
\nu f\left(\nu\right)=\xi \sqrt{\frac{2}{\pi}a \nu^{2}}\left[1+\left(a\nu^{2}\right)^{-p}\right]\exp\left[-a\nu^{2}/2\right] \,,
\label{scaling}
\end{equation}
where the normalization constant $\xi$ guarantees that $f$ is correctly normalized, {\it i.e.} $\int d \nu f\left(\nu\right)=1$. In all numerical applications of this expression presented here we will assume $p=0.3$ and $a=0.75$~\cite{Schmidt:2008tn}. 
\begin{figure}[t]
\centering
\includegraphics[width=0.6\textwidth,height=0.8\textwidth]{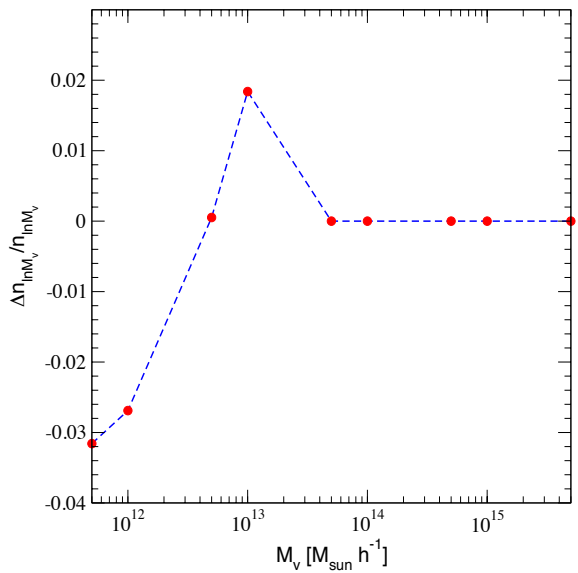}
\caption{Relative deviation from the $\Lambda$CDM prediction of the halo mass function obtained within the symmetron model. The dependence of $\delta_c$ on the virial mass has been included as explained in the text.  The red dots correspond to the masses for which we actually calculated the halo mass function, whereas the blue dashed curve connecting them has been drawn for illustrative purposes.}
\label{massfun}
\end{figure}

Contrary to the $\Lambda$CDM case, in the symmetron model the calculation of the halo mass function is complicated by the fact that $\delta_{c}$ depends on the initial radius of the collapsing structure, and therefore in turn on its virial mass. To handle this complication within the ST approach we have calculated $\delta_{c}(z=0)$ for a sample of virial masses spanning the mass range between $5\times10^{11} M_{\odot}~h^{-1}$ and $5\times10^{15}M_{\odot}~h^{-1}$, and then used these values of $\delta_c(z=0)$ to evaluate the mass function. For each mass in this range we have used the appropriate value of $\delta_c$. In Fig.~\ref{massfun} we show the halo mass function resulting from this procedure. Since the differences with respect to the $\Lambda$CDM case are at the few percent level, rather than showing the halo mass function itself, we plot the relative difference between the halo mass function of the symmetron model and the corresponding quantity calculated for the $\Lambda$CDM. The red dots correspond to the masses for which we actually calculated the halo mass function, whereas the blue dashed curve connecting them has been drawn for illustrative purposes. As one can clearly see from this figure, for virial masses larger than about $10^{13} M_{\odot}~h^{-1}$, the halo mass function of the symmetron model tends towards the one of the $\Lambda$CDM. The reason is that in the large virial mass limit the two models differ for their linear growth function only (as we can see from Fig.~\ref{Fig1}, if we increase the value of the initial radius, the sphere collapses at redshift closer to $\Lambda$CDM: for values of $R_{i}>5$ Mpc h$^{-1}$, the sphere collapses at the same redshift of $\Lambda$CDM). This quantity cancels in the ratio which defines the variable $\nu$, which thereby depends in both cases on the same initial conditions only.  
\begin{figure}[t]
\centering
\includegraphics[width=0.6\textwidth,height=0.8\textwidth]{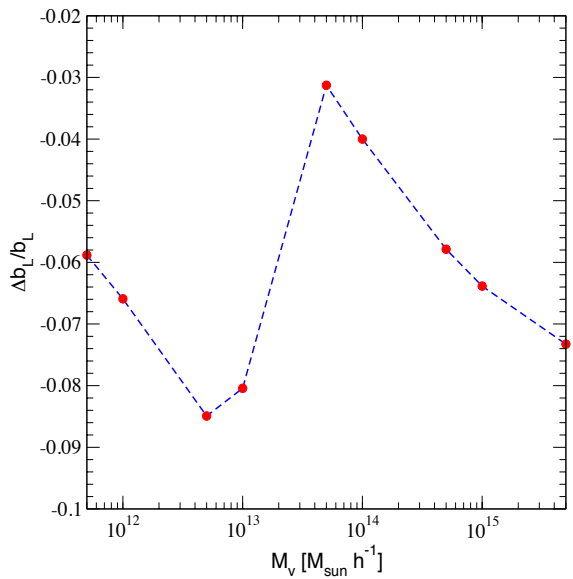}
\caption{Relative deviation from the $\Lambda$CDM prediction of the linear bias of the symmetron model. Since $b_L$  explicitly depends on $\delta_c$ (and not only through the variable $\nu$), contrary to the case of the mass function, also at large virial masses we observe percent deviations of our model from the $\Lambda$CDM prediction.}
\label{bL}
\end{figure}

We conclude this section focusing on the dark matter halo bias. Collapsed dark matter halos are biased tracers of the underlying dark matter distribution. This bias can be quantified comparing the halo-mass cross power spectrum with the matter power spectrum. Within the ST framework the linear bias ({\it i.e.} the bias in the limit $k\rightarrow 0$) takes the following form~\cite{Sheth:1999mn} 
\begin{equation}
b_{L}\left(M_{v}\right)=1+\frac{a\nu^{2}-1}{\delta_{c}}+\frac{2p}{\delta_{c}\left[1+\left(a\nu^{2}\right)^{p}\right]}\,.
\label{eq:bL}
\end{equation}
In Fig.~\ref{bL} we show the predictions of our model and compare them with standard $\Lambda$CDM expectations. Also in the case of the linear bias we employ the same approach outlined above to account for the dependence of $\delta_c$ on the viral mass. Since $b_L$  explicitly depends on $\delta_c$ (and not only through the variable $\nu$), also at large virial masses we observe percent deviations of our model from the $\Lambda$CDM prediction. The feature in the plot at about $10^{13} M_{\odot}~h^{-1}$ corresponds in fact to the transition between a regime where the variable $\nu$ coincides in the two scenarios to a regime in which $\nu$ is different in the two cases.

\section{Conclusions}
In this paper we have studied the formation of dark matter halos in the symmetron model, where a scalar field metrically coupled to all matter species alters the standard growth of cosmic overdensities. The dynamics of the symmetron is controlled by a $Z_2$ symmetry (under which this scalar is odd) whose breaking generates a new long-range interaction of gravitational strength.

We initially focused on single spherically symmetric overdensities whose evolution departs from the background expansion before recombination. We followed the time evolution of these objects generalizing the spherical collapse model to include all relevant physical effects related to the new long-range force mediated by the symmetron field. We identify two distinct scenarios, depending on the initial radius of the collapsing halo. For small initial overdensities, the evolution of the symmetron field inside the forming dark matter halo adiabatically follows during all the phases of the spherical collapse the one of the background scalar field. This implies that in this ``thick-shell scenario'' the formation of a dark matter halo is affected by the symmetron mediated force from the time of the $Z_2$ symmetry breaking until the dark matter halo has formed. In this case dark matter halos tend to collapse earlier compared to the $\Lambda$CDM case. In a second scenario, corresponding to large initial overdensities,  a transition between the previously described thick-shell regime and the ``thin-shell regime'' takes place, and  the symmetron force is screened in the interior of the collapsing halo before collapse. In this case dark matter halos tend to form later compared to the thick-shell scenario where instead the symmetron force was active until the end of the spherical collapse. Also in this case, however, halos collapse earlier than in the $\Lambda$CDM case where the additional long-range scalar force was not present at all. 

In the second part of this work we have instead analyzed the statistics of the dark matter halos formed within the symmetron model, focusing on the calculation of the halo mass function and of the linear bias for the two scenarios identified during the first part of this study. In both cases we compute the differences of the halo mass function and the linear bias obtained within the present realization of the symmetron model and the standard $\Lambda$CDM results. We find that  the relative difference between the halo mass function of our model and the one of the $\Lambda$CDM is typically of a few percent for halos smaller than  $O(10^{13} M_{\odot}~h^{-1})$ and vanishes for heavier halos. The deviation of the halo bias from the $\Lambda$CDM prediction  can be as high as $10\%$, and does not vanish for large masses.

The halo mass function enters the calculation of various observables of interest for cosmology and astroparticle physics, including the expected $\gamma$-ray signal induced by dark matter annihilations in extra-galactic halos~\cite{Ullio:2002pj}. Though the departure from the $\Lambda$CDM predictions found in this work seem very hard to identify at present, further investigations of the non-linear regime of the overdensity evolution based on dedicated N-body simulations might find additional features which would allow to better disentangle the halo mass function of the symmetron model from the one of the $\Lambda$CDM.

\section*{Acknowledgments}
M. Pietroni and R. Catena acknowledge partial support from the  European Union FP7  ITN INVISIBLES (Marie Curie Actions, PITN- GA-2011- 289442).

\end{document}